\documentclass[iop,apj]{emulateapj}
\usepackage{amsmath,amssymb, amsthm,amstext}
\usepackage{natbib}
\usepackage{graphicx}
\usepackage{color}
\usepackage{array}
\usepackage{bm}
\usepackage[breaklinks,colorlinks,citecolor=blue]{hyperref}

\def\nn{\nonumber}

\def\ct{\cos\theta}
\def\sst{\sin^2\theta}

\def\Ar{A_{\phi,r}}
\def\Arr{A_{\phi,rr}}
\def\Am{A_{\phi,\mu}}
\def\Amm{A_{\phi,\mu\mu}}

\def\be{\begin{equation}}
\def\ee{\end{equation}}
\def\ben{\begin{eqnarray}}
\def\een{\end{eqnarray}}
\def\WH{\Omega_{\rm H}}
\def\A0{A_{\phi}^{(0)}}

\def\AB{A_\phi^{\rm H}}

\def\Alfven{Alfv\`en}

\begin{document}
\title{Analytic properties of force-free jets in the Kerr spacetime -- III:\\
uniform field solution}

\slugcomment{Submitted to \apj}

\author{Zhen Pan}
\affil{Department of Physics, University of California, One Shields Avenue, Davis, CA 95616, USA;
\ {\textcolor{blue}{zhpan@ucdavis.edu}}}

\author{Cong Yu}
\affil{Yunnan Observatories, Chinese Academy of Sciences, Kunming 650011, China; \ {\textcolor{blue}{cyu@ynao.ac.cn}}\\
Key Laboratory for the Structure and Evolution of Celestial Objects, Chinese Academy of Sciences, Kunming 650011, China;}

\author{Lei Huang}
\affil{Key Laboratory for Research in Galaxies and Cosmology, Shanghai Astronomical Observatory, \\
Chinese Academy of Sciences, Shanghai, 200030, China; \ {\textcolor{blue}{muduri@shao.ac.cn}} }

\shorttitle{Analytic properties of force-free jets in the Kerr spacetime }
\shortauthors{Z. Pan, C. Yu and L. Huang}

\begin{abstract}
The structure of steady axisymmetric force-free magnetosphere of a Kerr black hole (BH) is governed
by a second-order partial differential equation of $A_\phi$ depending on two ``free" functions
$\Omega(A_\phi)$ and $I(A_\phi)$, where $A_\phi$ is the $\phi$ component of the vector
potential of the electromagnetic field, $\Omega$ is the angular velocity of the magnetic
field lines and $I$ is the poloidal electric current.
In this paper, we investigate the solution uniqueness.
Taking asymptotically uniform field as an example, analytic studies imply that
there are infinitely many solutions approaching uniform field at infinity,
while only a unique one is found in general relativistic magnetohydrodynamic simulations.
To settle down the disagreement, we reinvestigate the structure of the governing equation
and numerically solve it with given constraint condition and boundary condition.
We find that the constraint condition
(field lines smoothly crossing the light surface (LS))
and boundary conditions at horizon and at infinity
are connected via radiation conditions at horizon and at infinity, rather than being independent.
With appropriate constraint condition and boundary condition, we numerically solve the governing equation
and find a unique solution. Contrary to naive expectation,
our numerical solution yields a discontinuity in the angular velocity of the field lines
and a current sheet along the last field line crossing the event horizon.
We also briefly discuss the applicability of the perturbation approach to solving the governing equation.
\end{abstract}


\bigskip\bigskip

\section{Introduction}
The Blandford-Znajek (BZ) mechanism \citep{Blandford1977} is believed to be one of most efficient ways
to extract rotation energy from spinning black holes (BHs), which operates in BH
systems on all mass scales, from the stellar-mass BHs of gamma ray bursts to the
supermassive BHs of active galactic nuclei. In the past decade, we have gained
better understanding of the BZ mechanism from general relativistic
magnetohydrodynamic (GRMHD) simulations \citep[e.g.][]{Komissarov2001, Komissarov2004e,
Komissarov2004, Komissarov2005, Semenov2004, McKinney2004, McKinney2005, McKinney2007a,
McKinney2007b, Komissarov2007, Tchekhovskoy2008, Tchekhovskoy2010, Tchekhovskoy2011,
Palenzuela2011, Alic2012, Tchekhovskoy2012, Penna2013, McKinney2013}, numerical
solutions \citep[e.g.][]{Fendt1997, Udensky2004, Udensky2005, Palenzuela2010,
Contopoulos2013, Nathanail2014} and analytic perturbation solutions \citep[e.g.][]{Tanabe2008,
Beskin2013, Pan2014, Pan2015b, Pan2015, Pan2016, Gralla2014, Gralla2015, Gralla2016a,
Yang2015, Penna2015} to the steady axisymmetric force-free electrodynamics in the Kerr
spacetime.\footnote{A few families of exact solutions \citep[e.g.][]{Menon2005, Menon2007,
Menon2011, Brennan2013, Menon2015, Compere2016} to the equations of force-free electrodynamics
in the Kerr spacetime  have been found in the past decade. But these solutions
have various limitations, e.g., not allowing energy extraction from the BH,
being electrically dominated instead of magnetically dominated, lacking clear
physical interpretation, or being time-dependent and not axisymmetric, which
make it difficult to compare these exact solutions with simulations and
numerical solutions.} Various studies converge to a common picture of how the BZ mechanism operates:
The spinning BH distorts the poloidal magnetic field $\bm{B_{\rm P}}$, and induces the poloidal electric field $\bm{E_{\rm P}}$
and toroidal magnetic field $\bm{B_{\rm T}}$,
which generate an outward Poynting flux $\bm{E_{\rm P}
\times B_{\rm T} }$  along the magnetic field lines threading the spinning BH.
The rotation energy of the spinning BHs is extracted in the form of Poynting flux
\citep{Komissarov2009, Beskin2010}.

To step further, it is natural to ask whether
these different approaches give qualitatively and quantitatively consistent
descriptions of the BH magnetosphere structure, e.g.,
the topology of magnetic fields, the electric current distributions,
the angular velocities of the magnetic field lines
and the energy extraction rates. The answer is {\em yes} and {\em no}. The axisymmetric, steady-state,
force-free magnetosphere around  Kerr BHs is governed by the
general relativistic Grad-Shafranov (GS) equation. For
the simplest magnetic field configuration, split monopole
field, both analytic \citep{Pan2015b} and numerical solutions \citep{Nathanail2014}
reproduce the simulated angular velocity of field lines $\Omega$,
poloidal electric current $I$ and energy extraction rate $\dot E$ to high precision
\citep{Tchekhovskoy2010}. But for the asymptotically
uniform field, different approaches do not even reach a consensus on the solution uniqueness.
Time-dependent simulations \citep[e.g.][]{Komissarov2005, Komissarov2007, Yang2015} seem to converge to
a unique solution. Previous analytic studies \citep{Beskin2013, Pan2014, Gralla2016a} seem to find a unique
perturbation solution which roughly agrees with GRMHD simulations.
But in this paper, we will show there are actually many of them
due to the superposition of monopole component (and other possible components).
According to the argument of \citet{Nathanail2014},
solving the GS equation is actually an eigenvalue
problem, with two eigenvalues $\Omega(A_\phi)$ and $I(A_\phi)$ to be determined
by requiring field lines to smoothly cross the light surfaces (LSs).
For common field configurations, there exists usually two LSs, sufficing to determine two eigenvalues.
With only one LS for the uniform field configuration and one more boundary condition,
\citet{Nathanail2014} numerically found a unique solution, which however shows distinctive features from
previous GRMHD simulations \citep{Komissarov2005}.

How to explain the relationship between the unique solution
and the infinitely many possible candidates, and the discrepancy between
previous numerical solution and GRMHD simulations?
\footnote{It is definitely worthwhile examining the solution uniqueness problem of
the BZ mechanism, considering its significance in the modern relativistic astrophysics.
For example, rotating BHs are believed to be described by Kerr solution, but there is
no solid evidence for it. Some efforts have been done to detect possible deviation from Kerr solution
via BZ mechanism powered jet emission \citep[e.g.][]{Bambi2012b, Bambi2012, Bambi2015, Pei2016}.
A prerequisite for these BZ applications is its solution uniqueness.}
Does the plasma inertia make a difference? The force-free condition is assumed in
both analytic and numerical solutions, but the inertia cannot be completely
ignored in simulations. Taking account of the plasma inertia, \citet{Takahashi1990}
proposed the so-called MHD Penrose process, where the plasma particles within the
ergosphere are projected onto negative-energy orbits by magnetic field and
eventually are captured by the central BH. As a result,  \Alfven\ waves are
generated along the magnetic field lines, and BH rotation energy is carried away
by these \Alfven\ waves. \citet{Koide2002} and \citet{Koide2003} found the
MHD Penrose process was operating in GRMHD simulations (see \citealp[e.g.][]{Lasota2014,
Koide2014,Toma2014,Kojima2015,Toma2016}  for recent discussions on this issue).
If the MHD Penrose process is the dominant energy extraction process, the unique
solution found in simulations actually describes the MHD Penrose process instead
of the BZ mechanism. However, later simulations showed that the MHD Penrose process
is only a transient state, after which the \Alfven\ waves decay, the system settles
down into a steady state, and the BZ mechanism takes over \citep[e.g.][]{Komissarov2005}.
Therefore the plasma inertia seems to make little difference after the system settles
down into the steady state.

Another possible explanation is that, among all these mathematically possible
solutions, only the one found simulations is stable.
\citet{Yang2014} and \citet{Yang2015} analyzed the stability of these solutions,
and no unstable mode was found at order $O(a)$, where $a$ is the dimensionless
BH spin. Therefore modes can be unstable with a growth rate at $\sim O(a^2)$
at most.\footnote{The stability problem of general force-free jets is a complicated story.
Early studies implied that the force-free jets are vulnerable to various instabilities
\citep[e.g.][]{Begelman1998,Lyubarskii1999, Li2000, Wang2004}
while later studies showed that these instabilities are strongly suppressed by
field rotation, poloidal field curvature, etc
\citep[e.g.][and references therein]{Tomimatsu2001, McKinney2009a, Narayan2009g}.}
But the relevant timescale is
much longer than the  transient time scales observed in simulations.  Therefore
they concluded that the selection rule unlikely comes from instability.

In this paper, we show that the uniform field solution is
unique as strongly implied by previous GRMHD simulations
and pointed out by \citet{Nathanail2014}.
Following the algorithm proposed by \citet{Contopoulos2013} and \citet{Nathanail2014},
we numerically find a unique combination of $\Omega$ and $I$,
ensuring both smooth field lines across the LS and uniform field at infinity.
Contrary to \citet{Nathanail2014}, our numerical solution yields a discontinuity in the angular velocity of
field lines and a current sheet along the last field line crossing the event horizon,
which are features found in previous simulations.

We also investigate the applicability of analytic perturbation approach to the GS equation,
which relies on a fixed unperturbed solution and priorly known asymptotic behavior of magnetic field.
Analytic approach breaks down if any of the two factors is violated.
Both of the two are satisfied for monopole field in Kerr spacetime,
therefore we see the perfect match between high-order perturbation solutions,
and results from simulations and numerical solutions.
But for the uniform field, the unperturbed background field is not fixed due to
the superposition of the monopole component, therefore the perturbation approach
cannot predict a unique solution.

The paper is organized as follows. In Section \ref{sec:basic},
we summarize the basic equations governing the steady axisymmetric force-free magnetospheres.
In Section \ref{sec:sketch}, we clarify the relation between constraint conditions, radiation
conditions and boundary conditions; and our numerical method to solve the GS equation.
We apply the perturbation approach on the uniform field problem
and clarify the applicability of analytic perturbation approach in Section \ref{sec:pert}.
Summary and Discussions are given in Section \ref{sec:disc}. In Appendix, we present a robust solver
for the horizon regularity condition and its implication for the existence of electric current.

\section{basic equations}
\label{sec:basic}
In the force-free approximation, electromagnetic energy greatly exceeds that of matter.
Consequently, the force-free magnetospheres is governed by energy
conservation equation of electromagnetic field, or
conventionally called as the GS equation. In the Kerr spacetime,
the axisymmetric and steady
GS equation is written as \citep{Pan2014}
\ben
\label{eq:GS}
&-&\Omega \left[(\sqrt{-g}F^{tr})_{,r} +
(\sqrt{-g}F^{t\theta})_{,\theta} \right] + F_{r\theta}I'(A_\phi) \nn\\
&&+ \left[(\sqrt{-g}F^{\phi r})_{,r} +
(\sqrt{-g}F^{\phi\theta})_{,\theta} \right] = 0 \ ,
\een
which expands as
\citep[see also e.g.][in slightly different forms]{Contopoulos2013, Nathanail2014, Pan2016}
\ben
\label{eq:GSe}
&&\phantom{+}
 \left[\frac{\beta}{\Sigma}\Omega^2 \sst
-\frac{4ra}{\Sigma}\Omega \sst
-\left(1-\frac{2r}{\Sigma}\right)\right] \Arr  \nn\\
&&
+\left[\frac{\beta}{\Sigma}\Omega^2 \sst
-\frac{4ra}{\Sigma}\Omega \sst
-\left(1-\frac{2r}{\Sigma}\right)\right] \frac{\sst}{\Delta } \Amm \nn \\
&&
+ \left[  \Omega^2 \sst\left(\frac{\beta}{\Sigma}\right)_{,r}
-\Omega\sst\left(\frac{4ra}{\Sigma}\right)_{,r}
+\left(\frac{2r}{\Sigma}\right)_{,r}\right] \Ar \nn\\
&&
+\left[ \Omega^2\left( \frac{\beta\sst}{\Sigma} \right)_{,\mu}
-\Omega\left(\frac{4ra\sst}{\Sigma}\right)_{,\mu}
+\left(\frac{2r}{\Sigma}\right)_{,\mu}\right]\frac{\sst}{\Delta}\Am \nn \\
&&
+ \left[ \left(\frac{\beta}{\Sigma}\right) \Omega-\frac{2ra}{\Sigma}\right]
\sst\Omega'\left(\Ar^2 + \frac{\sst}{\Delta}\Am^2\right) \nn\\
&&
- \frac{\Sigma}{\Delta} II' = 0 \ ,
\een
where $\Sigma = r^2 + a^2 \mu^2$, $\Delta = r^2 -2r + a^2$, $\beta = \Delta\Sigma + 2r(r^2 + a^2)$,
$\mu\equiv\ct$ and  the primes designate  derivatives with respect to $A_\phi$.
For clarity, we may write the GS equation in a more illustrating form
\ben
\label{eq:GSg}
&&\phantom{+}
 \left[\Arr + \frac{\sst}{\Delta}\Amm \right]  \mathcal K(r,\theta; \Omega )\nn \\
&&
+\left[\Ar \partial_r^\Omega  +  \frac{\sst}{\Delta}\Am \partial_\mu^\Omega\right] \mathcal K(r,\theta; \Omega ) \nn \\
&&
+ \frac{1}{2}\left[\Ar^2 + \frac{\sst}{\Delta}\Am^2\right]  \Omega' \partial_\Omega \mathcal K(r,\theta; \Omega )\nn \\
&&
- \frac{\Sigma}{\Delta}II' = 0 \ ,
\een
where $\mathcal K(r,\theta; \Omega )$ is the prefactor of $\Arr$ in Equation~(\ref{eq:GSe}),
$\partial_i^\Omega (i=r, \mu)$  denotes the partial derivative
with respect to coordinate $i$ with $\Omega$ fixed, and $\partial_\Omega$ is the derivative with
respect to $\Omega$. The GS equation written in this compact form manifests clear symmetry,
therefore is beneficial in various aspects.\footnote{
For example, in flat spacetime, the classical pulsar
equation \citep{Scharlemann1973} is recovered by plugging
$\mathcal K(r,\theta; \Omega ) = \Omega^2 r^2\sst - 1$ into Equation~(\ref{eq:GSg}).}

\section{the solution uniqueness problem}
\label{sec:sketch}


In this section, we first clarify all the constraint conditions the GS equation satisfies,
and their relation with boundary conditions at horizon and at infinity.
We find the constraint conditions and boundary conditions are not independent.
For a given $\Omega(A_\phi)$, we can numerically
find a $I(A_\phi)$ ensuring field lines smoothly crossing the LS, but the combination of
$\Omega(A_\phi)$ and $I(A_\phi)$ obtained this way usually is in conflict with the uniform field boundary
condition at infinity. To be consistent this boundary condition, $\Omega(A_\phi)$ and $I(A_\phi)$ must
satisfy one more constraint. Then, we numerically find the unique combination of $\Omega(A_\phi)$ and $I(A_\phi)$
ensuring field lines smoothly cross the LS and being consistent with the boundary condition at infinity.
Finally, we compare our numerical solution with previous studies.

\subsection{Constraint Conditions and Boundary Conditions}
We want physically allowed solutions to be finite and smooth everywhere.
At LS where $\mathcal K = 0$, the second-order GS equation degrades to a first-order equation
\be
\label{eq:ls}
\begin{aligned}
& \left[\Ar \partial_r^\Omega  +  \frac{\sst}{\Delta}\Am \partial_\mu^\Omega\right] \mathcal K(r,\theta; \Omega ) \\
&+  \frac{1}{2}\left[\Ar^2 + \frac{\sst}{\Delta}\Am^2\right]  \Omega' \partial_\Omega \mathcal K(r,\theta; \Omega)
= \frac{\Sigma}{\Delta}II'.
\end{aligned}
\ee
Field lines smoothly crossing the LS must satisfy the above constraint,
which we call LS crossing constraint condition. At horizon and infinity,
the requirement of solution finiteness leads to the radiation conditions \citep[e.g.][]{Pan2016},
which read as,
\be
\label{eq:znajek}
    I = \frac{2r(\Omega-\WH)\sst}{\Sigma} \Am \Big|_{r=r_+},
\ee
and
\be
\label{eq:infty}
    I = -\Omega\sst \Am \Big|_{r\rightarrow \infty},
\ee
where $\WH$ is angular velocity of the central BH.

But the radiation conditions and boundary conditions are not independent.
For example, the radiation condition (\ref{eq:znajek}) uniquely determines the boundary values at horizon
if $\Omega$ and $I$ are specified, and we will use it as the inner boundary condition in our numerical calculation.
In the same way, the radiation condition (\ref{eq:infty})
uniquely determines the boundary values at infinity, if $\Omega$ and $I$ are specified;
or the radiation condition (\ref{eq:infty}) enforces a constraint on $\Omega$ and $I$,
if the boundary condition at infinity is given.
In our working example, the boundary condition at infinity
\be
\label{eq:rinfty}
A_\phi(r\rightarrow\infty) = r^2\sst
\ee
is given. Plugging it into the radiation condition (\ref{eq:infty}), we find
that $\Omega$ and $I$ must satisfy a new constraint \citep{Nathanail2014, Pan2014, Pan2016}
\be
\label{eq:constr}
I = 2\Omega A_\phi.
\ee
Note that conditions (\ref{eq:infty}, \ref{eq:rinfty}, \ref{eq:constr}) are not independent,
and we will use two of them (\ref{eq:rinfty}, \ref{eq:constr}) to close the GS equation.

Now we get two constraint conditions (\ref{eq:ls}, \ref{eq:constr}),
and two boundary conditions in the $r$ direction  (\ref{eq:znajek}, \ref{eq:rinfty}) ready,
(where the inner boundary condition (\ref{eq:znajek}) is nontrivial, see Appendix for details).
The next step is to specify proper boundary conditions in the $\mu$ direction.
According to the claim proved in paper II:
\noindent {\it ``In the steady axisymmetric force-free magnetosphere
around a Kerr BH, all magnetic field lines that cross the
infinite-redshift surface must intersect the event horizon",}
\footnote{The claim depends on the relation $I\propto \Omega$,
which can be derived from the radiation condition at infinity [Equation (\ref{eq:infty})]. But their are some debates about
whether the radiation condition holds for vertical field configurations.
As for the uniform field,
there is no disagreement on the relation $I\propto \Omega$ [Equation (\ref{eq:constr})],
though \citet{Pan2016} interpreted it as the result of radiation condition,
while \citet{Nathanail2014} interpreted differently.}
\noindent the possible field configuration in the steady state is shown in the Figure 1 of paper II.
Consequently, we write  boundary conditions in
the $\mu$ direction as follows,
\be
\label{eq:tbc}
\begin{aligned}
A_\phi(\mu = 1) & = 0 , \\
A_\phi\left(\mu = 0, r_+\leq r\leq 2\right) &= \AB ,\\
\Am\left(\mu = 0, r\geq 2\right) &= 0 ,
\end{aligned}
\ee
where the horizon enclosed magnetic flux $\AB$ is to be determined self-consistently.

\subsection{Numerical Method and Results}
The algorithm for numerically solving the GS equation was proposed by
\citet{Contopoulos2013} and was optimized by \citet{Nathanail2014}.
We slightly tailor their algorithm to accommodate the problem we are working on.
We define a new radial coordinate $R = r/(1+r)$, confine our computation domain
$R\times \mu$ in the region  $[R(r_+), 1]\times [0,1]$,
and implement a uniform $512\times 64$ grid.
The detailed numerical steps are as follows:

1. We choose some initial guess $A_\phi$, trial functions $\Omega$ and $I$ as follows,
\footnote{The convergent solution is independent of the trial field configuration or the grid resolution.
For example, we tested different initial field configurations $A_\phi = r^2\sin^2\theta  +\epsilon (1-\cos\theta)$,
different initial trial functions $\Omega$ and $I$, and different grid resolutions.}
\be
\begin{aligned}
    A_\phi &= r^2\sin^2\theta, \\
    \Omega &= \frac{\WH}{2}\cos\left(\frac{\pi}{2}\frac{A_\phi}{\AB}\right),\\
    I &= \WH A_\phi\cos\left(\frac{\pi}{2}\frac{A_\phi}{\AB}\right).
\end{aligned}
\ee

2. We evolve $A_\phi$ using relaxation method \citep{Press1987},
while this method does not work properly at the LS due to the vanishing second order derivatives.
Fortunately, the directional derivative of $A_\phi$ is known as a function of $II'$ there (see Equation (\ref{eq:ls})).
We instead update $A_\phi$ at the LS using neighborhood grid points and the directional derivative.
From the directional derivative and the grid points on the left/right side,
we obtain $A_\phi(r_{\rm ILS}^-)$/$A_\phi(r_{\rm ILS}^+)$.
Usually the two are not equal and field lines are broken here. To smooth the field lines,
we adjust $I(A_\phi)$ and update $A_\phi(r_{\rm ILS})$ as follows:
\be
\begin{aligned}
    II'_{\rm new}(A_{\phi, \rm new}) &= II'_{\rm old}(A_{\phi, \rm old}) \\
    &- 0.02(A_\phi(r_{\rm ILS}^+) - A_\phi(r_{\rm ILS}^-)) \ ,
\end{aligned}
\ee
where
\be
    A_{\phi, \rm new} = 0.5(A_\phi(r_{\rm ILS}^+) + A_\phi(r_{\rm ILS}^-)) \ .
\ee
Usually, $II'$ obtained is not very smooth.
To refrain possible numerical instabilities, we fit $II'(A_\phi)$ with a eighth-order polynomials.
In addition, $II'$ consists of two pieces: a regular piece determined as described above,
and a singular piece (the current sheet part)
\be
-\int_0^{\AB} II' dA_\phi \ \delta(A_\phi-\AB) \ .
\ee
In our compuation, we model the delta function as a parabola confined in
a finite interval $[\AB, \AB (1+\delta)]$ with $\delta = 0.1$ \citep[see e.g.][]{Gruzinov2005}.

3. Repeat step 2 for $10$ times, then update $\Omega(A_\phi)$ according to the constraint (\ref{eq:constr}).

\begin{figure*}
    \centering
\includegraphics[scale=0.45]{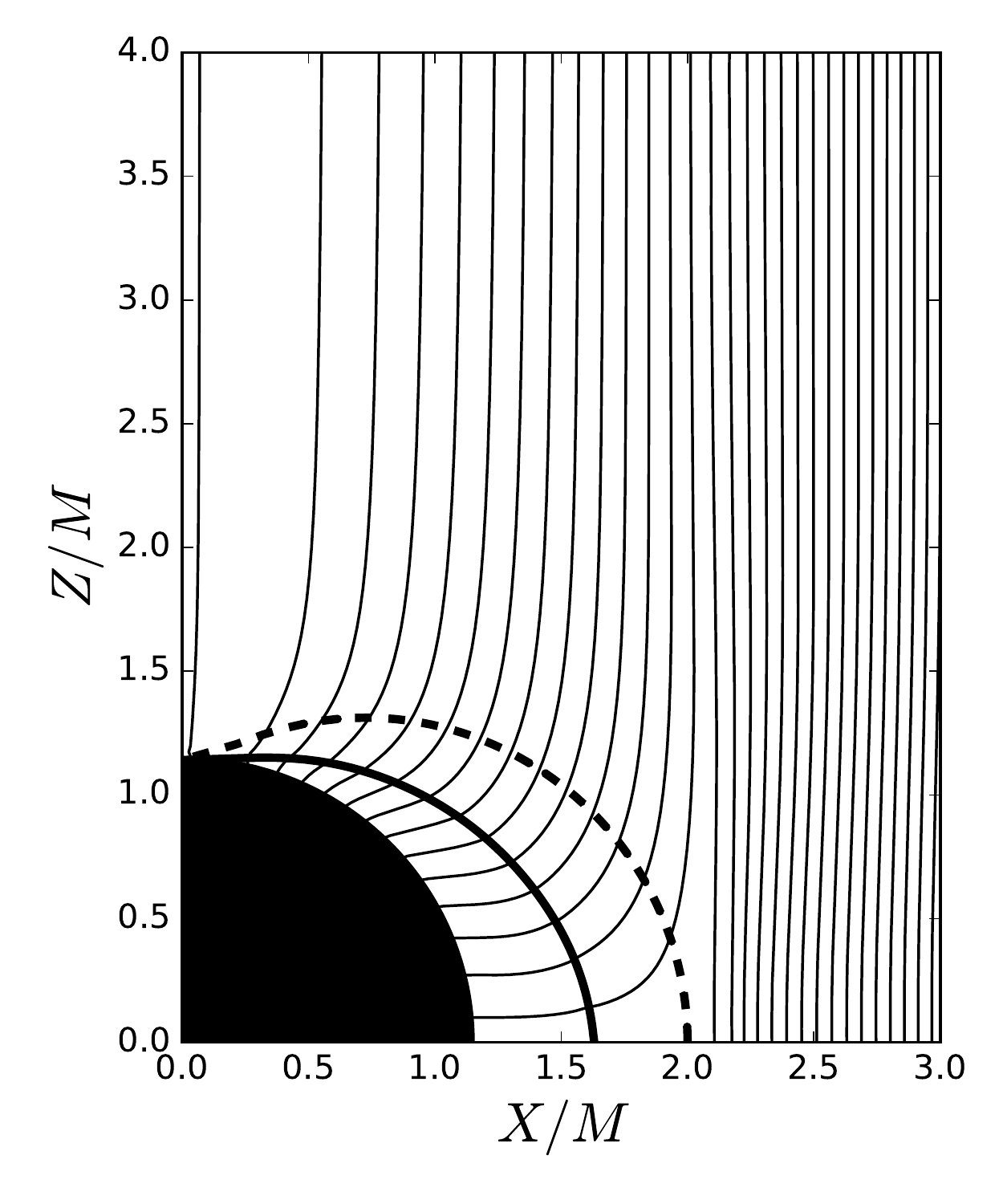}%
\includegraphics[scale=0.44]{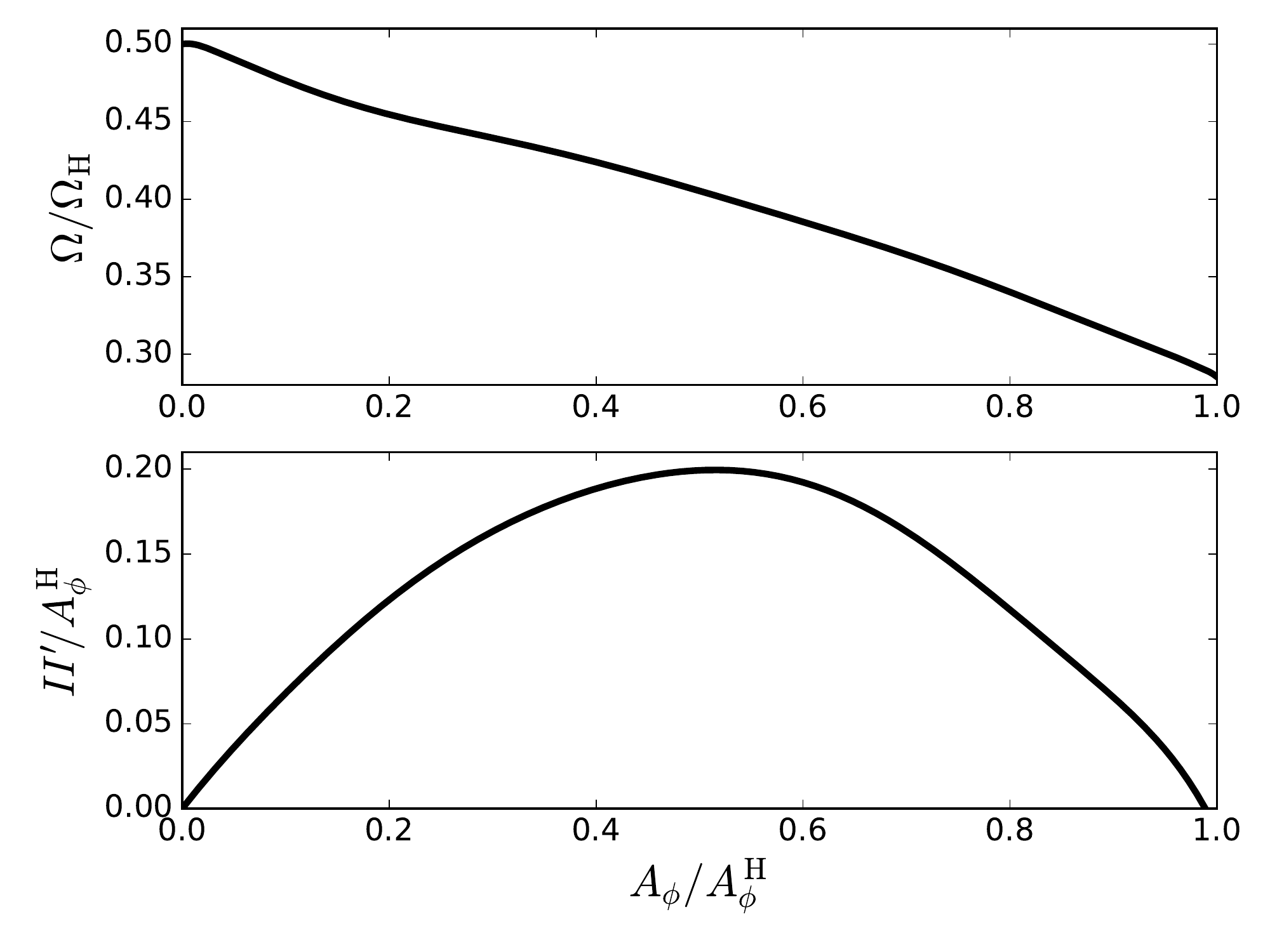}
\caption{Left Panel: The magnetic field lines around a Kerr BH with spin $a=0.99$,
where the inner(solid)/outer(dashed) curve is the ILS/IRS.
Right Panel: Functions $\Omega(A_\phi)$ and $II'(A_\phi)$ of the uniform field configuration,
where the latter is normalized by the amount of field flux crossing the event horizon $\AB$.}
\label{fig:lines}
\end{figure*}

We iterate the initial guess solution following the above steps until field lines smoothly
cross the LS and satisfy the boundary conditions.
The numerical results are shown in Figure \ref{fig:lines}.
In the left panel, we show the convergent field configuration
which as expected matches those of simulations \citep[e.g.][]{Komissarov2007}.
In the right panel, we show functions $\Omega(A_\phi)$ and $II'(A_\phi)$.
From the plot, we see that the angular velocity of the last field line crossing the event horizon
$\Omega(\AB)$ is not vanishing, i.e., $\Omega(\AB) \simeq 0.28\WH$,
while we expect the angular velocity of field lines not crossing the BH vanishes,
i.e., $\Omega(A_\phi > \AB) = 0$.

\subsection{Comparison with Previous Studies}

\citet{Nathanail2014} also studied BH magnetosphere structure of uniform field and concluded that
both $\Omega(A_\phi)$ and $I(A_\phi)$ must approach zero along the last field line crossing the event horizon,
and therefore the ILS coincides with the IRS along the equator,
and the electric current sheet does not appear. But as shown in the Appendix, the horizon regularity condition
requires the existence of current sheet along the equator. And our numerical solution shows that there is
a discontinuity in $\Omega(A_\phi)$ at $\AB$, and therefore the ILS lies between the event horizon
and the IRS, and there exists a current sheet.
The discrepancies here can be settled down by previous GRMHD simulations done by \citet{Komissarov2005},
where they observed a sharp transition in $\Omega(A_\phi)$ at $\AB$ and interpreted it as a discontinuity
smeared by numerical viscosity. It is worth noting that the discontinuity in $\Omega(A_\phi)$ does not lead to
any physical difficulties, e.g., the continuity of $\bm B^2- \bm E^2$ across the LS is not affected.

In addition, we do not explicitly show the BZ power of the uniform field configuration here,
because it is sensitive to the magnetic flux trapped by the event horizon $\AB$ which is boundary condition dependent.
In the real astrophysical environment, it is mainly determined by accretion process of the
central BH \citep[e.g.][]{Garofalo2009}.

\section{Applicability of Analytic Perturbation Approach}
\label{sec:pert}
In this section, we first recap the analytic perturbation approach to the GS equation,
then apply it to the uniform field problem \citep{Pan2014, Pan2015b} and explain
why this approach actually yields many solutions.
Finally, we discuss the applicability of the perturbation method.

We start with a unperturbed solution $A_0$ in the Schwarzschild spacetime,
\be
\label{eq:lgs}
\mathcal L A_0 =0,
\ee
where the operator
\be
\mathcal L\equiv\frac{\partial}{\partial
r}\left(1-\frac{2}{r}\right)\frac{\partial}{\partial r}
 +\frac{\sst}{r^2}\frac{\partial^2}{\partial \mu^2}.
\ee
For the corresponding Kerr metric solution $\{A_\phi, I(A_\phi), \Omega(A_\phi)\}$,
we assume $A_\phi|_{r\rightarrow\infty} = A_0$,
define $i = I|_{r\rightarrow\infty}, \omega = \Omega|_{r\rightarrow\infty}$,
and expand the solution to the leading order
\be
A_\phi = A_0 + a^2 A_2 , \quad \omega = a \omega_1 ,\quad i = a i_1.
\ee
Then we linearize the GS equation (\ref{eq:GSe}) as
\be
\mathcal L A_2(r,\theta) = S_2(r,\theta; i_1,\omega_1),
\ee
by only keeping leading order perturbation terms, where the source function $S_2$
depends on $i_1$ and $\omega_1$, which can be figured out from
the radiation conditions at  horizon and at infinity (\ref{eq:znajek}-\ref{eq:infty}).
The solution to the linearized GS equation is written as
\be
\begin{aligned}
A_2(r,\theta) =
\int\int dr_0  d\theta_0 \ S_2(r_0,\theta_0) G(r,\theta; r_0,\theta_0),
\end{aligned}
\ee
where $G(r,\theta; r_0,\theta_0)$ is the Green's function of operator $\mathcal L$ \citep{Petterson1974,Blandford1977}
\be
\mathcal LG(r,\theta; r_0,\theta_0) = \delta(r-r_0) \delta(\theta-\theta_0).
\ee
In this way, for a given Schwarzschild metric solution $A_0$, the corresponding
Kerr metric solution $\{A_\phi(a;,r,\theta), I(A_\phi),\Omega(A_\phi) \}$
is uniquely determined order by order.
Applying the method on the uniform field  problem $A_0=r^2\sst$, we find \citep{Beskin2013, Pan2014, Gralla2015},
\ben
\Omega = I = 0 && \quad (A_\phi > \AB), \\
\Omega = \frac{\WH}{2}\frac{\sqrt{1-A_\phi/\AB}}{1+\sqrt{1-A_\phi/\AB}}, && \ I=2\Omega A_\phi \quad (A_\phi < \AB).\nn
\een
where $\AB$ is the magnetic flux trapped by the event horizon
($\AB =4$ for the lowest-order perturbation solution, and generally depends on  BH spins and boundary conditions).

It seems that we have found the unique solution (at leading order) approaching uniform
field at infinity, but this is not the case.
The Schwarzschild spacetime GS equation (\ref{eq:lgs}) is linear.
Both uniform field and monopole field are its solutions, so do their linear combinations
\be
A_0(\epsilon) = r^2\sst + \epsilon (1-\ct),
\ee
where $\epsilon$ is some constant coefficient. The mixture of monopole component generates a
family of Schwarzschild metric solutions, $A_0(\epsilon)$, and all these solutions approach  uniform field at infinity.
For each solution, the corresponding Kerr metric solution $\{ A_\phi(\epsilon), I(\epsilon), \Omega(\epsilon) \}$
can be obtained using the above perturbation method.

To summarize, the perturbation method depends on two main ingredients:
the known asymptotic field behavior at infinity $A_\phi|_{r\rightarrow\infty}$
and the fixed underlying unperturbed field configuration $A_0$.
But for the uniform field problem,
there are many mathematically allowed unperturbed solutions
due to the additional monopole component.
That's why the perturbation approach cannot predict the unique solution.

\section{Summary and Discussions}
\label{sec:disc}
The GS equation is a second-order differential equation with two to-be-determined functions $\Omega$ and $I$.
Generally speaking, we need two constraint conditions to determine $\Omega$ and $I$,
two boundary conditions in the $r$ directions and two boundary conditions
in the $\mu$ direction to fix $A_\phi(r,\mu)$.
For asymptotically uniform field, we use constraint conditions (\ref{eq:ls},\ref{eq:constr}),
boundary conditions in the $r$ direction (\ref{eq:znajek},\ref{eq:rinfty})
and boundary conditions in the $\mu$ direction (\ref{eq:tbc}) to close the GS equation.
Our numerical solution of the uniform field yields a discontinuity in the $\Omega(A_\phi)$ at $\AB$,
therefore the ILS lies between the event horizon and the IRS, and there exists a current sheet along the last field
line crossing the event horizon (Figure \ref{fig:lines}), which is as expected from the horizon regularity condition.

Following the same logic, let us reexamine other two well studied field configurations:
monopole field in the Kerr spacetime and dipole field in the flat spacetime (classical pulsars).
For both field configurations, the number of LSs equals the number of to-be-determined functions $I$ and $\Omega$.
For monopole field in the Kerr spacetime, there are two LS crossing conditions and two radiation conditions.
The former two determine $\Omega$ and $I$, and the latter two determine the inner and the outer boundary.
Hence, there is no more freedom for specifying a boundary condition at infinity, i.e., we actually
do not know the solution at infinity before we really solve the GS equation.
Previous simulations and numerical solutions indeed confirmed
the asymptotic monopole field configuration $A_\phi \propto 1-\cos\theta$.
For pulsar dipole field, $\Omega$ is equal to the angular velocity of the central star
and $I$ is the only function to determine.
The only one LS uniquely determines $I$ \citep{Contopoulos1999},
and two radiation conditions automatically determine boundary conditions.
In the same way, there is no more freedom to impose a boundary
condition at infinity. And previous numerical studies found the field at infinity deviates from
$A_\phi \propto 1-\cos\theta$ \citep[e.g.][]{Gralla2016b}.

We also discuss the perturbation approach for solving the GS equation, whose
applicability depends on two main ingredients:
the known asymptotic field behavior at infinity $A_\phi|_{r\rightarrow\infty}$
and the fixed underlying unperturbed field configuration $A_0$.
For the monopole field, both of them are satisfied, therefore the perturbation approach
is applicable and the high-order perturbation solutions
show good match with results from simulations \citep{Pan2015b}.
For both uniform field in the Schwarzschild spacetime and dipole field surrounding static stars,
the superposition of monopole component (and possible other components) generates many unperturbed solutions,
as a result, the perturbation approach cannot predict
a unique uniform field solution in the Kerr spacetime or
a unique dipole field solution surrounding spinning stars.

\acknowledgements
We thank our referee Prof. I. Contopoulos for his insightful comments
on the validity of radiation condition for vertical field configurations and
on the possible difficulties arising form the discontinuity in the angular velocity of field lines.
We also thank A. Nathanail for clearly explaining his numerical algorithm.
C.Y. is grateful for the support by the National Natural Science Founda-
tion of China (grants 11173057, 11373064, 11521303),
Yunnan Natural Science Foundation (grants 2012FB187,
2014HB048), and the Youth Innovation Promotion Association, CAS.
Part of the computation was performed
at the HPC Center, Yunnan Observatories, CAS, China.
L.H. thanks the support by the National
Natural Science Foundation of China (grants 11203055).
This paper is supported in part by the Strategic Priority Research Program
``The Emergence of Cosmological Structures" of the Chinese Academy of Sciences,
grant No. XDB09000000 and Key Laboratory for Radio Astronomy, CAS.
This work made extensive use of the NASA Astrophysics Data System and
of the {\tt astro-ph} preprint archive at {\tt arXiv.org}.


\appendix
\label{sec:app}
For a given combination of $I(A_\phi)$ and $\Omega(A_\phi)$,
the horizon regularity condition (Equation (\ref{eq:znajek}))
uniquely determines the boundary condition at horizon $A_\phi(r= r_+, \mu)$.
We find its numerical solution is not trivial due to the nonlinearity.
To construct a robust solver, we first rewrite Equation (\ref{eq:znajek}) as
\be
    \mathcal I = \frac{2r(\Omega-\WH)\sst}{\Sigma} \mathcal A_{,\mu}\Big|_{r=r_+}
\ee
where we have defined two normalized variables, $\mathcal I = I / \AB$ and $\mathcal A = A_\phi / \AB$.
Here $\mathcal A $ runs from 0 to 1, and its values on boundaries are
$\mathcal A(\mu=0) = 1$ and $\mathcal A(\mu=1) = 0$.
Furthermore, we define $f(\mathcal A) \equiv \mathcal I/2(\WH - \Omega) $, and the above equation
is written in a variable separated form
\be
\frac{\mathcal A_{,\mu}}{f(\mathcal A)} = -\frac{r_+ \sst}{r_+^2 + a^2\mu^2},
\ee
which has a formal solution
\be
\label{eq:fs}
e^{\int_1^{\mathcal A(\mu)}\frac{d\mathcal A}{f(\mathcal A)}}
 = \frac{1-\mu}{1+\mu} \times e^{\frac{a^2}{r_+}\mu}.
\ee
In this form, numerically solving $\mathcal A(\mu)$ is stable.

Here we show a general property of force-free magnetospheres read out of the formal solution (\ref{eq:fs}):
the horizon condition requires the existence of a current sheet at the equator.
Enabling a non-singular solution, the integral $\int_1^{\mathcal A(\mu)}\frac{d\mathcal A}{f(\mathcal A)}$
must be finite, except at $\mathcal A = 0 (\mu = 1)$, where $f(\mathcal A) = 0$ due to vanishing $\mathcal I(\mu=1)$.
At $\mathcal A = 1 (\mu = 0)$, the finite integral requires nonzero $f(\mathcal A =1)$,
or quickly decreased $f(\mathcal A)$ (e.g., $\sim\sqrt{1-\mathcal A}$).
Usually $\Omega \leq \WH/2$, therefore $I(\AB)$ must be nonzero,
or quickly decrease to zero  (e.g., $\sim\sqrt{\AB-A_\phi}$), where the former is a current sheet at the equator,
and the later is a divergent current density (but weaker than the former).

\end{document}